\documentclass{article}  
\usepackage{breckenridge}
\newcommand \eg {{\it e.g.}}  
\frompage{000} \topage{000}                                              
\def\rightmark{Equilibrium in Heavy Ion Collisions}
\def\leftmark{V. Koch et al.}
 
\title{Equilibrium in Heavy Ion Collisions}

\authors{
{V. Koch and A. Majumder %
}\\[2.812mm]
{\normalsize
Nuclear Science Division\\
Lawrence Berkeley National Laboratory\\ 
Berkeley, CA 94720, USA \\[0.2ex] 
}}
 
\abstract{We discuss the question of equilibriation in heavy ion collisions
    and how it can be addressed in experiment}

 
\begin{document}
 
\maketitle
\setcounter{page}{1}

\section{Introduction}\label{intro}

The purpose of relativistic heavy ion collisions is to produce and study new
forms of matter. Therefore, one of the most basic requirements is that a
system close to thermal equilibrium is generated in these collisions. 
Considerable effort has gone into this question with an emphasis on the
description of particle abundances. Indeed, the hypothesis of a thermalized
system (thermal model) is very successful in reproducing and predicting
measured particle ratios of a wide range of energies and system sizes (for a
recent review see Ref. \cite{pbm}).

However, particle abundances provide only very limited information about 
the system and the actual (non-thermal) dynamics may simply be averaged out in
these observables. This is the underlying idea of the statistical model by
Fermi \cite{Fermi}, which assumes that phase space dominates simple (single
particle) observables of a multi-particle system. Indeed, the statistical
approach not only works for heavy ion systems but also for $e^+e^-$ or
proton-proton 
collisions. In the latter cases, there is hardly any matter produced which is 
in
equilibrium. 
Therefore, the success of the thermal model in reproducing the particle ratios
is a necessary, but not sufficient, condition for the existence of matter close
to equilibrium. 

\begin{figure}[htb]
\epsfxsize=0.6 \textwidth
 \centerline{\epsfbox{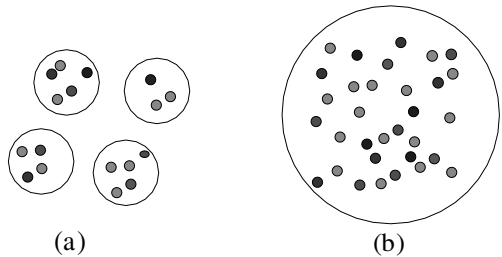}}   
 \caption{Individual nucleon-nucleon collisions (a) and nucleus-nucleus
 collision (b). }
\label{fig:1}
\end{figure}

We could for instance imagine that a heavy ion reaction is simply a collection
of many independent nucleon-nucleon collisions without any re-interaction of 
the secondary particles (Fig. \ref{fig:1}(a)). 
This system would still  look thermal in the
particle ratios, but hardly constitute matter (for a more
detailed discussion see Ref. \cite{koch_qm}). 
The other extreme, depicted in Fig. \ref{fig:1}(b), is matter equilibrated over
the entire volume of the system. Particle ratios alone cannot distinguish
between these two scenarios. 
But are there additional measurements, such as multi-particle correlations,
that allow us to distinguish between the two scenarios?

In the limit that each small system of Fig. \ref{fig:1}(a)
can be described in the {\em grand}-canonical approximation, there is no way
to distinguish between the two scenarios by simply measuring final state
particle yields and correlations. 
Even if the subsystems are separated by
arbitrary distances. In this limit,
the free energy of the system is just the sum of the free energies of any
subdivision, i.e. correlations are absent. This is for instance
the case for pions at high energies.
 
\begin{figure}[htb]
\epsfxsize=0.4 \textwidth
 \centerline{\epsfbox{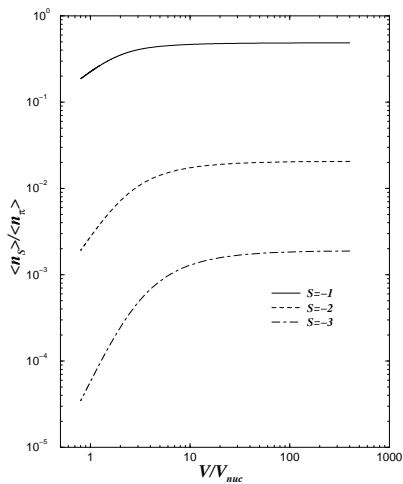}}   
 \caption{Abundance of strange particles as a function of the volume.}
\label{fig:2}
\end{figure}

However, additional conserved charges, such as strangeness introduce
correlations. In the scenario of Fig.\ref{fig:1}(a), strangeness needs to be
conserved in each small volume separately, thus reducing the yield of multiply
strange hadrons, such as the Omega baryon \cite{redlich-canonic}. This is
illustrated in Fig.\ref{fig:2}, where we show the abundance of 
strange particles (normalized to the pion number). 
Up to a volume of about 20 times that of a
nucleon-nucleon system, 
explicit strangeness conservation reduces the yield; most
dramatically for the $S=-3$ states.

The question is then: can we utilize the correlations introduced by the
strangeness conservation to determine the region over which strangeness is
equilibrated? 
To address this issue let us discuss the following model. 
Consider a system of total volume 
 $V_f$ divided into $p$ domains in which strangeness is explicitly
conserved. Suppose further that each of these
subsystems has the same temperature. This assumption is somewhat justified
since it is expected that kinetic equilibrium is reached faster than chemical
equilibrium. More importantly, in very high energy collisions of all species of 
`in states' (\eg, $e^+e^-$, $p,p$ and $A,A$), 
the freeze-out temperature assumes an 
almost universal value of $170\, \rm MeV$. Hence, it is reasonable to assume 
that each domain freezes out independently with a temperature $T \simeq 170 \,
\rm$. 
Furthermore, let us denote by $V_p = V_f/p$ the volume of each
domain. As we increase the number of domains or decrease the
domain volume{\bf,} the effect of local strangeness conservation becomes 
more prominent. As a result{\bf,} the number of
strange particles is reduced in the same fashion as depicted in
Fig.\ref{fig:2} (see also \cite{redlich-canonic}). 
However, for domain volumes 
\begin{equation}
V_p > 20 \, V_{nuc}{\bf,}
\end{equation}
the effect of strangeness conservation on the average particle number is small.
Here, $V_{nuc}$ denotes the volume of a nucleon-nucleon collision.
This is probably the case  at 
RHIC, where hardly any centrality dependence of the
strange particle yields is observed. At SPS energies the situation is
different. There,
NA57 \cite{manzari} measures a strong centrality dependence of the
$\Omega$-baryon yield, which is compatible with the rise of the $S=-3$ curve of
Fig.\ref{fig:2}(a) thus indicating a rather small strangeness equilibration
volume $V \leq 20 \, V_{nuc.}$ even for the most central collisions.
 \begin{figure}[htb]
 \epsfxsize=0.7 \textwidth
  \centerline{\epsfbox{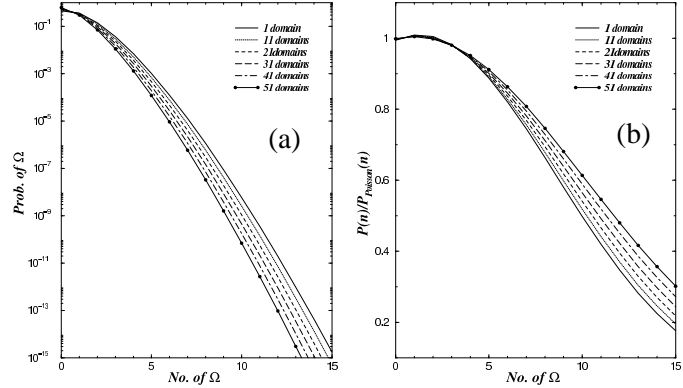}}   
  \caption{Probability distribution for $\Omega$-baryon for different numbers
  of domains (a). The right panel (b) shows the deviation from a Poisson distribution.}
 \label{fig:3}
 \end{figure}

So what can be done, if, as it is most likely the case at RHIC, the domain
volume is larger than $20 \, V_{nuc}$ and the sensitivity of the single
particle yield to the domain size becomes rather weak. 
The obvious idea is to
measure many particle coincidences. If the domain is sufficiently large, such 
that, 
the  production of three units of strangeness (such as an $\Omega$-baryon)
is not
affected by local strangeness conservation, then perhaps 
the production of 6 or
9 etc., units of strangeness is. This is demonstrated in Fig. \ref{fig:3}, where
we plot the probability distribution of $\Omega$-baryons for different number
of domains. Obviously, for $N_{\Omega} \geq 5$, the sensitivity on the
domain size is quite substantial. The probabilities to be measured are $P >
10^{-6}$ and are thus, in principle, accessible in a heavy ion experiment, where
of the order of $10^{6}$ events are being collected.   
 \begin{figure}[htb]
 \epsfxsize=0.4 \textwidth
  \centerline{\epsfbox{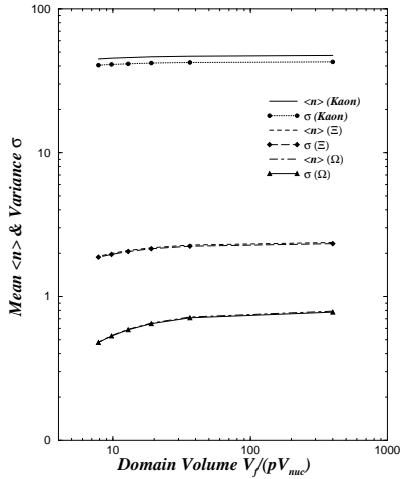}}   
  \caption{Mean and variance for different strange particles as a function of
  domain size.}
 \label{fig:4}
 \end{figure}

 In Fig.\ref{fig:3}(b) we
 show the ratio of the $\Omega$ probability distribution over that of a Poisson
 distribution. The Poisson distribution is fixed by the average value of the
 true distribution. We find that for $N_{\Omega} \leq 5$ the probabilities are
 essentially given by the Poisson distribution. Actually, the more
 domains we have, the closer we are to a Poisson distribution. At first
 sight, this is entirely counter intuitive. Many domains means small domain
 volume which in turn implies that the explicit strangeness conservation should
 be most relevant. Thus the system should deviate maximally from a
 grand-canonical description. At the same time we find a Poissonian behavior,
 which is 
 often considered an indication for a system in the 
 grand-canonical limit.

 The reason, we see Poisson like behavior in our case, is that for each small
 domain the probability to find an $\Omega$ is essentially binomial; either we
 find none or we find one. 
 \begin{eqnarray}
 P_{domain}(0) &=& 1 - \epsilon
 \nonumber \\
 P_{domain}(1) &=& \epsilon
 \nonumber \\
 P_{domain}(N>1) &\simeq& 0  
 \end{eqnarray}
 The probability to find more than one can be
 ignored. And having many domains, this binomial distribution turns in to a
 Poisson distribution as long as $N_{\Omega} \leq N_{domains}$.  
 Furthermore, since the probability 
 distribution falls off very rapidly (see Fig.\ref{fig:3}(a)), 
 two particle observables are determined by exactly the region where the
 distribution is very close to Poisson. As an amusing consequence, the
 variance is given by the mean (see Fig.\ref{fig:4}), 
 just as in the grand-canonical limit, although we are in a region where
 explicit strangeness conservation is essential.  
 Therefore, two particle observables do not provide new information about the
 size of the underlying domains. Quite to the contrary, they may lead to the
 wrong conclusion about the properties of the system at hand.

 \section{Conclusions}\label{concl}
 We have discussed the production of strange particles as a possible measure of
 the strangeness equilibrium volume. This volume will provide a lower bound on
 the equilibrium volume created in a heavy ion collision. We have shown, that
 the measurement of multiple $\Omega$-baryon final states with  
 $N_{\Omega} \simeq 5$
 can determine the size of this volume. We have further pointed out that two
 particle correlation will not provide new information
 compared to single particle measurements. 
 In conclusion, to establish the degree of equilibrium reached in these
 collisions requires a very precise and dedicated measurement of
 multi-particle final states. To which extent this can be achieved in present
 heavy ion experiments remains to be seen.

 \section*{Acknowledgments}
This work was supported in part by the Natural Sciences and Engineering 
Research Council of Canada and in part by the Director, 
Office of Science, Office of High Energy and Nuclear Physics, 
Division of Nuclear Physics, and by the Office of Basic Energy
Sciences, Division of Nuclear Sciences, of the U.S. Department of Energy 
under Contract No. DE-AC03-76SF00098.

\vfill\eject
\end{document}